\documentclass[showpacs,amsmath,amssymb,amsfonts]{revtex4}

\usepackage{enumerate}
\usepackage{graphicx}
\usepackage{dcolumn}
\usepackage[usenames]{color}
\usepackage{ulem} 
\providecommand{\av}[1]{\left\langle #1 \right\rangle} %

\def\p{\mathbf{p}}

\begin{document}
\title{Validity of the third law of thermodynamics for the Tsallis entropy}

\author{G. Baris Bagci}
\email{gbb0002@hotmail.com} \affiliation{Department of Materials Science and Nanotechnology Engineering, TOBB University of Economics and Technology, 06560 Ankara, Turkey}
\author{Thomas Oikonomou}
\affiliation{Department of Physics, Nazarbayev University, 53 Kabanbay Batyr Avenue, Astana 010000, Kazakhstan}


\begin{abstract}

Bento \textit{et al.} [Phys. Rev. E 91, 022105 (2015)] recently stated that the Tsallis entropy violates the third law of thermodynamics for $0 < q <1$ in the sub-additive regime. We first show that the division between the regimes $q < 1$ and $q > 1$ is already inherent in the fundamental incomplete structure of the deformed logarithms and exponentials underlying the Tsallis entropy. Then, we provide the complete deformed functions and show that the Tsallis entropy conforms to the third law of thermodynamics for both super-additive $q < 1$ and sub-additive $q > 1$ regimes. Finally, we remark that the Tsallis entropy does not require the use of escort-averaging scheme once it is expressed in terms of the complete deformed functions.

\end{abstract}


\maketitle

\section{introduction}

Statistical mechanics successfully provides the microscopic foundation for the macroscopic laws of thermodynamics. Nevertheless, the laws of thermodynamics are still fundamental, since statistical mechanics aims to recover them from the underlying atomic structure. Therefore, any generalization of statistical mechanics should satisfy the laws of thermodynamics to describe the macroscopic phenomena.

One such generalization is the Tsallis entropy \cite{Tsallis} whose goal is to explain the inverse power law distributions found in nature. In this context, it has found various applications such as black hole thermodynamics \cite{Cirto}, high energy physics \cite{Deppman}, and quantum information \cite{Rajagopal}. Despite these vast number of applications though, there are still some unsettled issues related to the Tsallis entropy at a fundamental level such as its continuum generalization \cite{Abe} and zeroth law of thermodynamics \cite{Van}.

One such issue is whether the Tsallis entropy conforms to the third law of thermodynamics as recently pointed out in Ref. \cite{Bento}. Interestingly, Bento \textit{et al.} have shown that the third law of thermodynamics is violated by the Tsallis entropy for the sub-additive regime $q < 1$. The third law is an important and essential test for any generalized entropy measure as emphasized in Ref. \cite{Bento}, since it should be satisfied for any Hamiltonian independent of modelling short-range or long-range interactions. Therefore, it is worth careful study and analysis \cite{Bento2,Bento3}.

Before proceeding further, we note that the demarcation of the intervals in the Tsallis formalism is also evident in the so-called finite bath scenario \cite{Bath}. Considering only the escort distributions used in Ref. \cite{Bento}, the distribution with $q < 1$ ($q >1$) corresponds to a physical system in contact with heat bath possessing a positive (negative) finite heat capacity \cite{Bath}. Moreover, these two regimes also differ from one another in their behaviours, namely, the distribution with $q < 1$ represents a sharp decay with a cut-off while the distribution with $q > 1$ is a genuine inverse power-law decay with fat tails. In other words, according to Ref. \cite{Bento}, the Tsallis entropy conforms to the third law of thermodynamics when it yields the power-law decay i.e. $q > 1$ whereas the same entropy expression violates the third law when it results in a sharp decay with a cut-off for $q < 1$.

We will show in the next section that this occurrence of the interval demarcation between $q < 1$ and $q > 1$ is due to the incomplete use of the deformed functions inherent in the Tsallis entropy right from the beginning. Then, we provide the solution to this issue. Section III revisits the third law of thermodynamics through the complete deformed function by which we show that the third law of thermodynamics is valid for both $q < 1$ and $q > 1$. Finally, the conclusion is presented in Section IV.

\section{Completing the Tsallis entropy}
We begin by the Boltzmann-Gibbs entropy which reads

\begin{equation}\label{Shannon}
S_{BG} = \av{\ln(1/p_\lambda)} = \av{-\ln(p_\lambda)} 
\end{equation}

\noindent where $p_{\lambda}$ is the probability of the $\lambda$th microstate and $\av{.}$ denotes linear averaging. This expression is possible since one simply has
\begin{equation}\label{log}
\ln(1/p_\lambda) + \ln(p_\lambda)   = 0.
\end{equation}
The Tsallis entropy \cite{Tsallis} reads
\begin{equation}\label{tsallis1}
S_q = \av{\ln_{q}(1/p_\lambda)}=\frac{\sum_{\lambda} p_{\lambda}^q-1}{1-q}
\end{equation}
where the deformed $q$-logarithm $\ln_q (x)$ is generally defined as
\begin{equation}\label{q-log-main}
\ln_q(x)=\frac{x^{1-q}-1}{1-q}.
\end{equation}
In other words, the Tsallis entropy completely relies on the linear averaging scheme exactly as the Boltzmann-Gibbs entropy. The sole difference between the two is that the Tsallis entropy makes use of the $q$-deformed logarithm whereas the Boltzmann-Gibbs entropy uses the ordinary (natural) logarithm. In the limit $q \rightarrow 1$, the $q$-deformed logarithm becomes the ordinary (natural) logarithm so that the Tsallis entropy reduces to the Boltzmann-Gibbs one.

Despite this resemblance in their fundamental structure, Eq. \eqref{log} is not satisfied by the $q$-deformed logarithms since
\begin{equation}\label{qlog}
\ln_{q}(1/p_\lambda) + \ln_{q}(p_\lambda) = \frac{p_\lambda^{1-q}+p_\lambda^{q-1}-2}{1-q} \neq 0. 
\end{equation}
The above relation indicates that the use of the deformed logarithms (and also their inverse deformed exponentials) is limited to certain values of the deformation parameter $q$ independent of the physical system under scrutiny. Therefore, it is not clear whether the calculated $q$ range in an arbitrary application of the non-extensive formalism is a result of the physical model or an artefact of the above relation inherent in its fundamental structure. The relation similar to Eq. \eqref{log} can be found as
\begin{equation}\label{qlog2}
\ln_{q}(1/p_\lambda) + \ln_{2-q}(p_\lambda) =  0, 
\end{equation}
as can be checked by using Eq. \eqref{q-log-main} \cite{Oikonomou}.

However, the question remains as to whether Eq. \eqref{qlog2} can be used for all $q$ values. In order to see this, we propose the following consistency check: we first consider the ordinary canonical distribution with the inverse temperature $\beta$ and energy $E_{\lambda}$ i.e. $p_\lambda \varpropto \exp \left( - \beta E_{\lambda} \right) $ apart from the well-behaving (i.e. it does not diverge) partition function. This expression shows that when $\beta\rightarrow+\infty$ then $p_\lambda$ tends to zero and vice versa in a unique manner. Accordingly, the following limit must be satisfied
\begin{equation}\label{limit1}
\lim_{p_{\lambda} \rightarrow 0} \ln(1/p_\lambda) 
=- \lim_{p_{\lambda} \rightarrow 0}\ln(p_\lambda) = +\infty, 
\end{equation}
\noindent due to the monotonicity of the logarithm as a necessary and sufficient condition that the canonical distribution $p_\lambda$ has to fulfil. Only through this limit, one can ensure that the condition $p_{\lambda} \rightarrow 0$ can be satisfied also at the level of the canonical distribution simultaneously warranting $\beta \rightarrow +\infty$ (see \cite{Oikonomou} for details).

Applying the consistency condition in Eq. \eqref{limit1} to the $q$-deformed logarithms given by Eq. \eqref{qlog2}, we see that   
\begin{equation}\label{limit2}
\lim_{p_{\lambda} \rightarrow 0} \ln_{q}(1/p_\lambda)=
-\lim_{p_{\lambda} \rightarrow 0} \ln_{2-q}(p_\lambda)=\quad
\begin{cases}
+\infty&,q\in(0,1]\\
\frac{1}{q-1}&,q\in[1,2)\\
\end{cases}
\end{equation}
%
%
%
The equation above show that the relation $\lim_{p_{\lambda} \rightarrow 0} \ln_{q}(1/p_\lambda) = -\lim_{p_{\lambda} \rightarrow 0} \ln_{2-q}(p_\lambda) =+ \infty$ is satisfied only for $q\in(0,1]$. Having observed the validity interval, we can write the Tsallis entropy as
\begin{equation}\label{tsallis2}
S_{q\in(0,1]} = \av{\ln_{q}(1/p_\lambda)}= \av{- \ln_{2-q}(p_\lambda)} =\frac{\sum_{\lambda} p_{\lambda}^q-1}{1-q},
\end{equation}
which is only valid for $q\in(0,1]$ as the index denotes.

In order to obtain the Tsallis entropy for the interval $q\in[1,2)$, we make the transformation $p_{\lambda} \rightarrow \frac{1}{p_{\lambda}}$ (or, equivalently, the transformation $q \rightarrow (2-q)$) in Eq. \eqref{qlog2} so that we obtain  
\begin{equation}\label{qlog3}
\ln_{q}(p_\lambda) + \ln_{2-q}(1/p_\lambda) =  0. 
\end{equation}
The consistency condition in Eq. \eqref{limit1} applied to the $q$-deformed logarithms above yields
\begin{equation}\label{limit3}
-\lim_{p_{\lambda} \rightarrow 0} \ln_{q}(p_\lambda)=
\lim_{p_{\lambda} \rightarrow 0} \ln_{2-q}(1/p_\lambda)=\quad
\begin{cases}
+\infty&,q\in[1,2)\\
\frac{1}{1-q}&,q\in(0,1]\\
\end{cases}
\end{equation}
%
%
%
so that the Tsallis entropy for the interval $q\in[1,2)$ reads 
\begin{equation}\label{tsallis3}
S_{q\in[1,2)} = \av{ \ln_{2-q}(1/p_\lambda)} =\av{- \ln_{q}(p_\lambda)}=  \frac{\sum_\lambda p_{\lambda}^{2-q}-1}{q-1}.
\end{equation}

Before proceeding with the third law of thermodynamics, the explanation of a very subtle issue is in order: the maximization of the Tsallis entropy $S_{q\in(0,1]}$ in Eq. \eqref{tsallis2} subject to the ordinary internal energy $U = \sum_{\lambda} p_{\lambda} E_{\lambda}$ (together with the normalization condition) yields the ordinary $q$-exponential distribution 
\begin{equation}\label{ord}
p_{\lambda} \left( E_{\lambda} \right) =  \left[  1+(1-q)\beta \left(E_{\lambda}-U \right)        \right]^{\frac{1}{q-1}}
\end{equation}  
which is valid only for $q\in(0,1]$, since this equilibrium probability distribution is obtained through the use of $S_{q\in(0,1]}$ in Eq. \eqref{tsallis2} \cite{Oikonomou}. On the other hand, the maximization of the Tsallis entropy $S_{q\in[1,2)}$ in Eq. \eqref{tsallis3} again subject to the same constraint $U = \sum_{\lambda} p_{\lambda} E_{\lambda}$ yields the escort $q$-exponential distribution 
\begin{equation}\label{esc}
p_{\lambda} \left( E_{\lambda} \right)=  \left[  1+(q-1)\beta \left(E_{\lambda}-U \right)        \right]^{\frac{1}{1-q}}
\end{equation}  
which is valid only for $q\in[1,2)$, since this equilibrium probability distribution is obtained through the use of $S_{q\in[1,2)}$ in Eq. \eqref{tsallis3} \cite{Oikonomou}. In passing, we also note another strong evidence for the validity intervals of the distributions in Eqs. \eqref{ord} and \eqref{esc} above: $p_{\lambda} \left( E_{\lambda} \right)$ in Eq. \eqref{ord} is indeed normalizable only for $q\in(0,1]$ while $p_{\lambda} \left( E_{\lambda} \right)$ in Eq. \eqref{esc} is normalizable only for $q\in[1,2)$ in terms of the argument $E_{\lambda}$.

It is important to note that the escort distribution in Eq. \eqref{esc} is generally obtained in the literature by maximizing the expression $S_{q\in(0,1]}$ in Eq. \eqref{tsallis2} but by subjecting it to the constraint $U = \frac{\sum_{\lambda} p_{\lambda}^{q} E_{\lambda}}{\sum_{\lambda} p_{\lambda}^{q}}$ together with the normalization condition. Then, deemed necessary by the application at hand, the transformation $(2-q)$ is used to obtain the ordinary $q$-exponential distribution in Eq. \eqref{ord}. This kind of formalism has three drawbacks: one is the necessity of the justification of the constraint $U = \frac{\sum_{\lambda} p_{\lambda}^{q} E_{\lambda}}{\sum_{\lambda} p_{\lambda}^{q}}$. The second drawback is the inconsistency in using the constraint $U = \frac{\sum_{\lambda} p_{\lambda}^{q} E_{\lambda}}{\sum_{\lambda} p_{\lambda}^{q}}$, which is not formed by a linear averaging scheme through $p_\lambda$ otherwise inherent in the expression of the Tsallis entropy (see either Eq. \eqref{tsallis2} or Eq. \eqref{tsallis3}) \cite{Aver}. The third drawback is the justification of the $(2-q)$ transformation used in an \textit{ad hoc} manner out of necessity for some applications \cite{Tsallis}.

The complete Tsallis formalism consists of $S_{q\in(0,1]}$ and $S_{q\in[1,2)}$ together with the ordinary normalization and internal energy constraints, namely, $\sum_{\lambda} p_{\lambda} = 1$ and $U = \sum_{\lambda} p_{\lambda} E_{\lambda}$, respectively. It shows that the constraint $U = \frac{\sum_{\lambda} p_{\lambda}^{q} E_{\lambda}}{\sum_{\lambda} p_{\lambda}^{q}}$ is not essential and the ordinary internal energy constraint $U = \sum_{\lambda} p_{\lambda} E_{\lambda}$ is sufficient for all admissible $q$ values. It also clarifies the origin of the otherwise unjustified $(2-q)$ transformation through the relations given by Eqs. \eqref{qlog2} and \eqref{qlog3}. 
A more compact representation of the complete Tsallis formalism combining $S_{q\in(0,1]}$ and $S_{q\in[1,2)}$ in a single expression is provided through the use of   the complete deformed $q$-logarithms and exponentials (see also Ref. \cite{Oikonomou} for details)
\begin{equation}\label{list}
\ln_q(x):=\frac{x^{1-q}-1}{1-q}
		\bigg|_{
		x\in(0,1]\wedge q\in[1,2)}^{
		x\in[1,\infty)\wedge q\in (0,1]}
\,,\qquad
\exp_q(x):=\Big[1+(1-q)x\Big]^{\frac{1}{1-q}}
		\bigg|_{
		x\in(-\infty,0]\wedge q\in[1,2)}^{
		x\in[0,\infty)\wedge q\in(0,1]}\,,
\end{equation}
which is used together with its $(2-q)$ counterpart.

\newpage

\section{The Third law of Thermodynamics}
%

In order to assess whether the Tsallis entropy conforms to the third law of thermodynamics, following Ref. \cite{Bento}, we begin by assuming that microscopic energies $E_{\lambda}$ where $\lambda = 0,1,...,N$ are ordered (without degeneracy). The probability $p_{\lambda = 0}$ is the one of ground state. Assuming moreover the normalization $\sum_{\lambda} p_{\lambda}=1$, we write $p_0=1-\sum_{n} p_{n}$ with $n=1,...,N$. Note that the relation $\frac{\partial f\left( p_{0}\right) }{\partial p_{n}} = -\frac{\partial f\left( p_{0}\right) }{\partial p_{0}}$ holds where $f$ is any function of the micro-probabilities. The third law scenario dictates that the ground state is fully populated whereas all the other states are empty so that we set $p_0$ to unity and all the remaining $p_n$'s to zero. The inverse temperature due to the $n$th energy level $\beta_{n}$ is given by
\begin{eqnarray}\label{temp}
\beta_{n}= \frac{\partial S}  {\partial p_{n}} \left({\frac{\partial U}{\partial p_{n}}}\right)^{-1}.
\end{eqnarray}
where the total inverse temperature obeys  $\beta = \sum_{n} \beta_{n}$ \cite{Bento}. The third law test then checks whether this inverse temperature $\beta_{n}$ approaches infinity when $\left\lbrace p_{n}\right\rbrace \to0$ as $p_{0}=1$ 
as a result of only the lowest energy level being occupied while all the other levels are empty \cite{Bento}.

We now check whether the Tsallis entropy conforms to the third law of thermodynamics for the interval ${q\in(0,1]}$. For this interval, the Tsallis entropy is given by Eq. \eqref{tsallis2}. Therefore, we obtain  
\begin{equation}\label{thirda}
\frac{\partial S_{q\in (0,1]}}{\partial p_n}=-q\Big[\ln_{2-q}(p_n)-\ln_{2-q}(p_0)\Big].\
\end{equation}
Since the internal energy is given by $U = \sum_{\lambda} p_{\lambda} E_{\lambda}$, we have  
\begin{equation}\label{energya}
\frac{\partial U}{\partial p_n}=E_n-E_0.
\end{equation}
The substitution of the two equations above into Eq. \eqref{temp} yields
\begin{equation}\label{thirdb}
\beta_n^{q\in (0,1]}=-q\lim_{\substack{p_n\rightarrow0\\p_0\rightarrow1}}
		\frac{\ln_{2-q}(p_n)-\ln_{2-q}(p_0)}{E_n-E_0}
=-q\lim_{p_n\rightarrow 0}\frac{\ln_{2-q}(p_n)}{E_n-E_0}=+\infty
\end{equation}
which is valid only for the interval $q\in (0,1]$.

Concerning the interval $q\in [1,2)$, we use the Tsallis entropy in Eq. \eqref{tsallis3} so that 
\begin{equation}\label{thirdc}
\frac{\partial S_{q\in [1,2)}}{\partial p_n}=(q-2)\Big[\ln_q(p_n)-\ln_q(p_0)\Big].
\end{equation}
Noting that Eq. \eqref{energya} for the internal energy expression is again valid, we see that 
\begin{equation}\label{thirdd}
\beta_n^{q\in [1,2)}=(q-2)\lim_{\substack{p_n\rightarrow0\\p_0\rightarrow1}}
		\frac{\ln_q(p_n)-\ln_q(p_0)}{E_n-E_0}
=(q-2)\lim_{p_n\rightarrow 0}\frac{\ln_q(p_n)}{E_n-E_0}=+\infty
\end{equation}
which is valid only for the interval $q\in [1,2)$. Combining Eqs. \eqref{thirdb} and \eqref{thirdd}, we see that the Tsallis entropy conforms to the third law of thermodynamics for all possible values of the non-additivity index $q$.

\section{Conclusion}

The Tsallis escort distributions exhibit different behaviours in different regimes of the non-additivity index $q$. For the interval $q < 1$, they yield a sharp decay whereas they result in the power-law decay for $q > 1$ \cite{Bath}. A similar kind of demarcation of the relevant $q$ intervals also becomes apparent when one tests the Tsallis entropy through the third law of thermodynamics: the Tsallis entropy violates the third law for $q < 1$ while it conforms to it for $q > 1$ \cite{Bento}.

This demarcation issue can be resolved if one carefully considers the domain of applicability of the deformed functions such as $q$-logarithms and/or exponentials. By completing these deformed functions through the justified $(2-q)$ transformation (see Eqs. \eqref{limit1}, \eqref{limit2} and \eqref{limit3}  for example), we showed that the Tsallis entropy does conform to the third law of thermodynamics for both $q < 1$ and $q > 1$.

Another interesting result of the above analysis is that one does not need the escort averaging scheme at all. In this regard, we show that the role of the escort averaging is only to act as a bridge between the ordinary linear averaging suitable for $0 < q \leq 1$ and the escort distributions suitable for $1 \leq q < 2$ \cite{Oikonomou}. Once one uses the complete formalism of the deformed functions, the escort averaging becomes redundant for the third law of thermodynamics.

Finally, we would like to consider the experimental evidences put forth in Ref. \cite{Bento} for the validity of the third law only in the interval $1 \leq q \leq 2$. These evidences are cold atoms in dissipative optical lattices \cite{Renzoni} and the velocities of particles in driven-dissipative two-dimensional dusty plasma \cite{Liu}. 
A close inspection (see Eq. (1) in both Ref. \cite{Renzoni} and Ref. \cite{Liu}) shows that both these experimental results consider the distribution in Eq. \eqref{esc} (and not the one in Eq. \eqref{ord}), which is only valid for $q \geq 1$ \cite{note}. 
Note that the same reasoning also applies to the very important recent experimental validation of the scaling law in confined granular media with $q\geq1$ [see Eq. (3) in Ref. \cite{Combe}]
Therefore, the fact that these physical systems (with their distributions normalizable only for $q \geq 1$) conforms to the third law only for $q \geq 1$ does not imply that the third law is not valid for the Tsallis entropy for $q < 1$.


\begin{thebibliography}{00}
\bibitem{note} The case could not have been otherwise, since the distribution in these experimental results are normalizable only in the interval $1 \leq q < 2$. Ref. \cite{Renzoni} explicitly states this normalizability condition below Eq. (1). Note that the upper limit of the interval in Ref. \cite{Renzoni} is not $2$ but $3$ due to the squared dependence $x^2$ of the argument of the $q$-Gaussian.
\end{thebibliography}
\end{document}